# PC Cluster Machine Equipped with Low-Latency Communication Software

Motohiko Tanaka

Coordinated Research Center, Institute for Fusion Science
Toki 509-5292, Japan
Email: mtanaka@nifs.ac.jp    http://dphysique.nifs.ac.jp/

## Abstract

A high performance Beowulf (PC cluster) machine installed with Linux operating system and MPI (Message Passing Interface) for interprocessor communications has been constructed using Gigabit Ethernet and the communication software GAMMA (Genoa Active Message Machine), instead of the standard TCP/IP protocol. Fast C/Fortran compilers have been exploited with the GAMMA communication libraries. This method has eliminated large communication overhead of TCP/IP and resulted in significant increase in the computational performance of real application programs including the first-principle molecular dynamics simulation code. Scalability of application programs on the number of processors and reliability over a long period of time (days) have also been confirmed for the GAMMA communications.

## 1. To Reduce Large Latency of TCP/IP

An old-day dream of "Constructing a super-computer with personal computers" became reality in 1990's North America [1,2] where a large number of personal computers (PC) with distributed memories were connected with each other by network. Each unit of the PC cluster machine, sometimes called a Beowulf cluster machine, has a scalar-type processing unit and its own memory. Thus, high performance computation comparable to a supercomputer is made possible at orders of magnitude lower cost. In this country, PC cluster machines were introduced in various universities [3,4]. Our group also made a series of PC cluster machines with Pentium III, Xeon, and Pentium 4 together with the Linux operating system and MPI (Message Passing Interface) [5] to study physics and chemistry of advanced ionic materials [6,7].

In order to realize high performance parallel (cluster) computers, all the three features of fast processors (cpu), high data bandwidth to/from the processors, and interprocessor communications are required. It is customary to use special hardware to realize fast computations and communications, which are most advanced on state-of-art vector/parallel supercomputers and also as designated chips for workstations. Some examples are Grape chips specialized for calculating gravity and Coulomb forces [8], and Myrinet for fast communications [9]. However, these hardware are more expensive than a PC cluster itself, and their use in the PC cluster is not without a question in terms of the total cost of the system.

Despite of the popularity of PC cluster machines because of their ultra high cost-performances, the presence of large communication overhead due to latency of the TCP/IP protocol has not widely been recognized. In fact, resource information by the "top" command shows that cpus are not 99% busy all the time. Thus, replacing the fast Ethernet (100Mbits/s) with gigabit Ethernet does not speed up the computation of real application programs. As depicted in Table 1, for example, the timing of the program that includes frequent MPI communications with the same number of either Pentium 4 or RISC processors makes the communication problem very clear. For the former with the TCP/IP communications, the wallclock time is significantly larger than the cpu time; their ratio is almost 1.4 with four processors of Pentium 4 in the first-principle molecular dynamics simulation. But, the communication overhead is very small for the latter.

In the following sections of this article, we describe the method of installing and using the low-latency communication software GAMMA (Genoa Active Message Machine) [10], which is a non TCP/IP communications and works on fast/gigabit Ethernet. Also, we first show the way to use fast C/Fortran compilers that are not included in the standard Linux distributions. The introduction of these new elements result in removal of the large communication overhead of the TCP/IP communications and significant increase in computational abilities of the PC cluster machine for real application programs, only at small additional cost for the second Ethernet. Accuracy and reliability of the GAMMA communication system over a day's span are also confirmed for the real application program.

The SCore project [11] generally yields good performance of high asymptotic bandwidth, but more latency on the gigabit Ethernet compared with GAMMA.

## 2. Utilization of GAMMA and Fast C/Fortran Compilers

In this section, we describe the method of installing the GAMMA and MPI programs [10] to a PC cluster machine. We also show the way how to use the fast C/Fortran compilers that are not contained in the Linux distributions with the

GAMMA system. In principle, the system is based on the active message mechanism [12] that enables direct communications between the application program and the network interface while bypassing the operating system. (Detailed procedures of the installation are found in the GAMMA installation manual, and in Appendices here; updated information at URL of Ref.[7].)

Nice features of the present method is that all the software including the Linux operating system and GAMMA software are either free or low priced, and that the hardware including processors, gigabit Ethernet cards and switching hubs are reasonably priced commercial products. On the other hand, for obtaining high efficiency of communications, it is recommended to have dual networks; the first network is a gigabit Ethernet for the GAMMA data transmissions, and the second one is a TCP/IP network for the NFS file system and general administration purposes, as shown in Fig. 1.

The programs to be installed are GAMMA, (modified) MPI-1.1 [5] and the MPI interface for GAMMA, all of which are downloaded from the same site [10]. Some remarks are: (i) not all the NIC (network interface card) are supported for fast/gigabit Ethernet, and (ii) the proper Linux kernel (currently version 2.4.21) must be chosen due to assembler of the GNU gcc compiler. If necessary, the Linux kernel should be downloaded and upgraded.

### 2(a) Linux Kernel Upgrade
The Linux kernel is an independent part of the Linux distributions, and can be updated by following the procedures given in Appendix A. Since upgrading the kernel requires special carefulness and some trials and errors, old data and configurations should be backed up in a safe place for emergency. During the upgrade, the existing kernel should never be removed, which may be used under the dual-boot environment for emergency recovery.

### 2(b) Installation of GAMMA
Installing procedures of GAMMA are easier than the Linux kernel upgrade. Configure and compile the source code, and write it onto the kernel (Appendix B). In the configuration, the flow control option should be turned on (see Sec.3). After the installation of GAMMA software, some environmental settings are necessary to initiate communications (Appendix C). This includes creating the list of nodes that participate in the GAMMA communications. Then, one should test the communication by invoking the pingpong program; if everything has properly been established, the communication speed is shown against specified size of transmitted data.

### 2(c) Installation of MPI/GAMMA
To utilize the GAMMA communications from application programs, the MPI and interface to the GAMMA programs need to be installed (Appendix D). For consistency with the GAMMA libraries, the GNU C/Fortran included in the Linux distributions should be used for this compilation. Note that, if messages "No rule to make….Stop." has appeared, the instruction should be followed and include files be copied as directed (unfortunately, compilation does not stop here). The GAMMA communication system guarantees correctness of the transmitted data for the blocking-mode communications. On the other hand, it is essential to reset communication status by invoking the "gammaresetall" command to maintain node synchronization when tasks have failed by any reasons.

### 2(d) Fast C/Fortran Compilers
For getting high computational efficiency and/or using the Fortran90 standards, the use of commercial compilers are recommended instead of the Linux's GNU C/Fortran compilers. It is very important here to arrange for consistency between the MPI/GAMMA libraries and the execution binaries of user's application program

(Appendix E). Most essential here is to attach two trailing underscores to function /subroutine names, which is the GNU C/Fortran standards. Due to this reason, all the mathematics libraries including BLAS and LAPACK, and their extensions to parallel computation BLACS and SCALAPACK [13] need to be recompiled.

There is one important warning about the logical operations with the (all)reduce function, which always returns .false. irrespectively of the input values on some compilers. This is due to the definition (the use of integers) of the logical true and false values which differs from that of the GNU compilers [14]. This is a very hard trap to be located if not informed [15].

## 3. Performance of MPI/GAMMA System

Let us examine how the computational speed increases in actual application programs when the MPI/GAMMA communication system is combined with fast C/Fortran compilers.

First, the data transmission speed is shown in Fig.2 as a function of the transmitted data size [16]. This timing refers to a point-to-point communication measured by the pingpong program included in the GAMMA program. The data transmission speed is defined by the transmitted data size divided by the time spent on them. A 3Com996 gigabit network interface card is used with Pentium 4 processors (3.0GHz). The transmission speed 0.6Mbits/s for one byte data corresponds to the latency 15micro sec. This small latency is quite advantageous for real application programs that exchange small data frequently over the conventional TCP/IP protocol whose latency is about 100micro sec (this point will be again mentioned in Table 1). The transmission speed increases with the data size, and saturates around $10^5$ bytes. Asymptotic band width for the present environment (no special optimization) is 706Mbits/s which amounts to 70% of the gigabit Ethernet.

The best values shown on the GAMMA homepage [10] are the followings. For Myrinet (1.28Gbits/s) on the BIP platform, the latency and asymptotic bandwidth are 4.3micro sec and 1005Mbits/s, respectively, and are 8.5micro sec and 976Mbits/s for GAMMA + gigabit Ethernet NIC Netgear GA621, respectively.

Next, the computational efficiency of the GAMMA communication system is examined for the real research application: density-functional first-principle (quantum mechanical) molecular dynamics code Siesta [17]. This is a tight-binding ab initio code adopting atomic orbital bases, whose most demanding operations are diagonalization of the density matrix under parallel algorithm.

As the test case, we choose a liquid system consisting of 180 atoms of carbon, nitrogen and hydrogen and a proton in slab geometry [18]. Table 1 shows that the communication overhead decreases dramatically when the TCP/IP communication (top row) is replaced by the GAMMA communication with the flow control enabled (middle row); the overhead time 26sec for TCP/IP is reduced to 0.1sec for GAMMA, and consequently, the wallclock time for one SCF cycle of the Siesta run decreases from 93sec for TCP/IP to 66sec for GAMMA. The time spent on the processing units is quite similar for these cases. If the flow control is not enabled during the GAMMA communications, the wallclock time degrades substantially due to retransmission of the data lost by collision. This should apply to usual application programs that we use in our research.

For the reference, timing of the same Siesta code for the typical RISC machine IBM Power 4, operated under parallel circumstances (slightly slower than SGI Altix3000), is shown at the bottom row of Table 1. It tells us that the PC cluster machine made of Pentium 4 processors via the GAMMA communications is nearly equivalent in computational speed with the same number of RISC machines having half the clock speed. Small overhead times are common to these machines.

Thus, by allocating a reasonable number (4-8) of processors for a single job, the PC cluster machine can be as powerful as a vector/parallel supercomputer. In some special case, the computational speed of the PC cluster exceeds that of the supercomputer with the same number of processors. This happens for the tight-binding first-principle (quantum mechanical) molecular dynamics simulations.

Figure 3 shows how the computational speed scales with the number of processors when the Siesta code is used on the GAMMA system. Here, the ordinate is the inverse of the wallclock time for one SCF cycle of the run (181 atoms, slab geometry) normalized to that of the uni-processor case. The computational speed increases nearly linearly up to four processors, and improves gradually beyond that. The computation speed by GAMMA communications is about 1.5 times that by TCP/IP [19].

We estimate the portion of non-parallelizable part P in the simulation program (computational time based). The relative computational speed is approximated by Amdahl's law $1/[P+(1-P)/N]$, where N is the number of used processors. From this formula and Fig.4, one estimates P= 0.10 (10%) for GAMMA, and P= 0.23 (23%) for TCP/IP. Thus, it is important to use a low latency communication system for the computational speed to scale linearly with the number of processors ($P \ll 1$), since the communication overhead constitutes a non-parallelizable part.

## 4. Conclusion

In this article, it was shown that the large latency of the TCP/IP communication, which was a bottleneck for computational speedup of the PC cluster machine, can be removed by adopting the low latency communication software GAMMA which works via the gigabit Ethernet. Also, the method of using efficient C/Fortran compilers on this low latency communication system was described, with detailed procedures given in Appendices.

The timing results show that the PC cluster machine of Pentium 4 and the GAMMA system via the gigabit Ethernet is almost equivalent in the computational speed with the RISC machine (of half the clock speed) of the same number of processors. It was also found that the PC cluster machine is as useful as a vector/parallel supercomputer in some application programs that contain frequent inter-processor communications. This includes the tight-binding first-principle molecular dynamics simulations.

Finally, scalability and reliability of the GAMMA communication system have been tested by running the consecutive Siesta runs for a day's span. During these runs, communication error has not been detected. On the other hand, increase in occupied RAM memory has been observed, whose needs to be suppressed in the near future.

**Acknowledgments:**
The author thanks Dr. Giuseppe Ciaccio for his kind advices on the installation of the GAMMA system to his Beowulf cluster machine. He also thanks Dr.Y.Zempo for close collaboration on the construction of the cluster machine and the installation of the first-principle molecular dynamics code Siesta. The timing with IBM Power 4 was performed using the Minnesota University Supercomputing Institute. The present work was supported by the Grant-in-Aid No.16032217 (2003-2005) from the Japan Ministry of Education, Science and Culture.

**Appendices:**

The following procedures and numerical accuracies have been examined for the combination of Pentium 4 and Red Hat Linux 7.3. Note that the kernel upgrade and MPI/GAMMA program installation procedures require special carefulness and some efforts with trials and errors. Please make sufficient preparations including backup of existing data and configurations, and try these things at your own risk. The location of files may slightly differ among Linux distributions.

**Appendix A: Kernel Upgrade**

A kernel is the Linux core which is independent of the Linux distributions. The kernel that GAMMA currently requires is of the version 2.4.21 (will be 2.6.x this fall). To upgrade it, first download the kernel source code (from http://www.linux.or.jp/, for example). In the following procedures, # stands for a command prompt (waiting for your command input). You need to be a root to alter the system directories.

#mv linux-2.4.21.tar.gz /usr/src    (2.4.21 kernel is used)
#cd /usr/src
#rm -rf linux    (delete the old link only, and never delete the old kernel itself !)
#tar xvzf linux-2.4.21.tar.gz
#ln -s linux-2.4.21 linux    (establish a new link)
#cd /usr/src/linux
   Now edit the Makefile in this directory; uncomment the #export INSTALL_PATH= /boot
   (remove #). To reuse existing system configurations, copy the old configuration to .config:
#cp /usr/src/linux-(old)/configs/kernel-(old)-(arch).config /usr/src/linux/.config
   Here (old) should be a previous kernel version number, and (arch) is the architecture number
   like i386 or i686.

#make oldconfig    (skip all lines by an Enter key)
#make xconfig    (Use an X window to work in the interactive mode)
   Choose a proper "Processor Family" for your system, set "Symmetric multi-processing
   support" on. Choice items are extensive; help messages are shown by clicking on HELP
   located at the right-hand side of each line of the menu panel. You may use default values if you
   are not sure of them.
#make dep
#make clean    (always remove stale files)
#make bzImage    (build the kernel image)
   If you find errors in this step, some of your choices above were not right. Return to "make
   xconfig", and redo "make dep; clean; bzImage" steps until you see no errors. These steps may
   require trials and errors.    After done, you are ready to write the properly configured kernel image to your hard disk.

#mkdir /lib/modules/2.4.21

#/sbin/installkernel 2.4.21 arch/i386/boot/bzImage System.map

Now edit either /etc/lilo.conf or /etc/grub.conf if you use the LILO or GRUB boot loader. The label (LILO) or the number (GRUB) after "default=" specifies the new kernel to be booted automatically. It is essential that you retain the old kernel entry for emergency rescue.

Next, make modules for the new kernel:

#make modules

#make modules_install

#/sbin/depmod -a (write out module dependencies; make sure no errors appear)

# /sbin/mkinitrd -ifneeded /boot/initrd-2.4.21.img 2.4.21

#/sbin/lilo -v   (if LILO is used)   or   # /sbin/grub-install /dev/had (if GRUB)

In above, replace /dev/hda with /dev/sda if the boot disk is a SCSI drive.

After done, reboot your PC and choose the new kernel. Should you not find an X Window or Ethernet connection, you only need to add a video or NIC driver to the new kernel; download the source and compile it. However, if you encounter serious errors like "Kernel panic", you have to reboot the PC and work on the old kernel to correct errors. Go back to the beginning of the kernel upgrade; start with "#make clean" to delete stale files and dependencies. If you remake the modules, delete old modules by "#rm -rf   /lib/modules/2.4.21/*" first.

**Appendix B: Installing GAMMA**

Download from http://www.disi.unige.it/project/gamma/ the source of GAMMA, MPICH-1.1.2, and MPI/GAMMA. The GAMMA program is assumed to be put under the /usr/local directory. (Read the GAMMA Installation Manual located in the /doc directory.)

#cp gamma-(version).tar.gz /usr/local

#tar xvzf gamma-(version).tar.gz

#cd /usr/local/gamma

#./configure

Here configure GAMMA with a proper choice of NIC and mode of communications. In order to avoid data collisions during communication, set "Flow control" on. For the remote command shell, "rsh" is conveniently chosen; list up the host names that participate in GAMMA communications in /etc/hosts.equiv on all nodes, and enable rsh, rexec, rlogin daemons by "# /sbin/chkconfig rsh on" etc. Instead, you can choose ssh.

Then, compile GAMMA and place it into the Linux kernel.

#cd /usr/src/linux

#make xconfig   (work on an X window)

Deselect the TCP/IP driver that conflicts with the GAMMA driver.

```
#make dep
#cd /usr/local/gamma
#make    (compile GAMMA)
#make install
#cd /usr/src/linux
#make bzImage    (make a new kernel image that contains GAMMA)
#/sbin/installkernel 2.4.21 arch/i386/boot/bzImage System.map
```
   Edit /etc/grub.conf and delete one of duplicated boot entries. Set GAMMA environments (see Reference C). Then, reboot your PC to put the new configuration into effect..

   To make use of the two independent networks per PC, mount the home directory of the master node via TCP/IP network as NFS (network file system). This is required for a job to refer to the execution binary located on the master node.

**Appendix C: Environmental Settings for GAMMA**

Following settings are necessary for GAMMA to operate properly.

* When rsh is used for the remote shell, start the rshd, rlogind, rexecd daemons by
  "# /sbin/chkconfig rlogin on", etc.
* NFS mount the home directory of the master node on the slave nodes. To do this, list up in /etc/exports of the master node the directory names to be exported and the slave node names to which the directories are exported (to export /home to all nodes under 192.168.1 with read/write permissions, the entry is: /home 192.168.1.0/24(rw)). Then, start the service by "# /sbin/service nfs start". On the slave nodes, NSF mount the directory by "#mount master_node_name:directory_name local_name". Conveniently, add the NSF entry in the /etc/fstab of the slave nodes so that "#mount local_name" should work.

(1) In /etc/gamma.conf of all nodes, list up the pairs of the host name of eth0 NIC and its MAC address (12-digit hardware address, shown by "# /sbin/ifconfig"), one entry per line. For example, pc001 as a host name and 01:23:45:67:89:ab as MAC address should be paired as:
    pc001 0x01 0x23 0x45 0x67 0x89 0xab
(2) Add "export PWD" in .bash_profile of user's home directory.
(3) To boot GAMMA automatically, add "/usr/local/bin/gammagetconfig" in the file /etc/rc.d/rc.local    (may differ among Linux distributions).

Run the test program pingpong under the user mode (rsh is usually disabled under the root mode for security).

**Appendix D: Installing MPI/GAMMA**

Use mpich-1.1.2 specially modified to use with GAMMA.
```
#cp mpich-1.1.2.tar.gz mpigamma-(version).tar.gz /usr/local
```

#tar xvzf mpich-1.1.2.tar.gz    (after expansion, sub-directory mpich is created)

#tar xvzf mpigamma-(version).tar.gz

#cd /usr/local/mpich

Now configure MPI/GAMMA using the following script (make this script file executable; in the below, ¥ stands for a line continuation.)

./configure -cc=gcc -fc=f77 -cflags=-fomit-frame-pointer -optcc=-O3 ¥
-noromio -noc++ -nompe -prefix=/usr/local/lib ¥
-lib=/usr/lib/libgamma.a -device=gamma

#make    (do not execute "#make install")

[**Warning**] During this make, an error message "No rule to make target ….Stop." appears but compilation never stops ! (Strange errors while compiling or executing an application program using MPI/GAMMA very often originate here). To correct reference errors, copy include files to designated directories as required, and do make again. Specifically, the new directories are those with the name "linux" instead of "linux-2.4.21". Copy /usr/src/linux-2.4.21/include/linux/gamma/ libgamma.h and /usr/src/linux-2.4.21/include/asm/page.h to appropriate new directories.

After the make, libraries and include files are placed in /usr/local/mpich/build/LINUX/ gamma/lib and include directories.

**Appendix E: Fast C/Fortran Compilers with GAMMA**

\* First of all, copy mpif.h, mpi.h, mpi_errno.h and binding.h files in the /usr/local/ mpich/include directory to /usr/local/mpich/build/LINUX/gamma/include.   Make sure that the first columns of mpif.h are (!) instead of (c).

\* Consistency is required among the GAMMA libraries and application program execution binaries. Mathematics libraries need to be recompiled under the following conditions:

(i) Linux's GNU C/Fortran compilers attach two trailing underscores (__) after the subroutine /function names. This rule must be followed by other compilers,

(ii) Cite include files of GAMMA/mpich,

(iii Link farg.f   which ports different C/Fortran compilers [14].

(iv) An MPI routine mpi_allreduce.c must be modified and compiled to absorb the differences of the definitions for the logical constant .true. (=1 fog GNU compilers, and = -1 for PGI compilers) [15].

(v) Recompile the linear algebra libraries BLAS, LAPACK, and their extension to parallel computations BLACS, SCALAPACK [13] under the conditions (i) and (iv).

\* **Compilation script for Portland's pgf90 compiler to use with MPI /GAMMA**

pgf90 -o ax1.out -Msecond_underscore -Mvect (Fortran program name) ¥
-I/usr/local/mpich/build/LINUX/gamma/include ¥

farg.o 　-L/usr/lib/libgamma.a ¥

　　(SCALAPACK and BLACS libraries, if necessary) ¥

　　-L/usr/ local/mpich/build/LINUX/gamma/lib -lfmpich -lmpich ¥

　　/usr/local/BLAS/libblas.a /usr/local/LAPACK/liblapack.a

**\*Modified allreduce.c (located in /usr/local/mpich/src/coll)**

```
/*
 *  Special version to use with pgi/gcc   updated: 2004/12/03 mtanaka (NIFS)
 *
 *      pgi -> gcc (Allreduce) -> pgi
 *
 *  $Id: allreduce.c,v 1.2 1998/04/28 18:50:43 swider Exp $
 *
 *  (C) 1993 by Argonne National Laboratory and Mississipi State University.
 *      See COPYRIGHT in top-level directory.
 */

#include "mpiimpl.h"
#include "coll.h"
#include "mpiops.h"

/*@
MPI_Allreduce - Combines values from all processes and distribute the result
                back to all processes
Input Parameters:
+ sendbuf - starting address of send buffer (choice)
. count - number of elements in send buffer (integer)
. datatype - data type of elements of send buffer (handle)
. op - operation (handle)
- comm - communicator (handle)

Output Parameter:
. recvbuf - starting address of receive buffer (choice)

.N fortran
.N collops
.N Errors
.N MPI_ERR_BUFFER
.N MPI_ERR_COUNT
.N MPI_ERR_TYPE
.N MPI_ERR_OP
.N MPI_ERR_COMM
@*/
int MPI_Allreduce ( sendbuf, recvbuf, count, datatype, op, comm )
void            *sendbuf;
void            *recvbuf;
int              count;
MPI_Datatype     datatype;
MPI_Op           op;
MPI_Comm         comm;
{
    int mpi_errno = MPI_SUCCESS;
    struct MPIR_COMMUNICATOR *comm_ptr;
    struct MPIR_DATATYPE     *dtype_ptr;
```

```c
    MPIR_ERROR_DECL;
    static char myname[] = "MPI_ALLREDUCE";

    TR_PUSH(myname);

    comm_ptr = MPIR_GET_COMM_PTR(comm);
    MPIR_TEST_MPI_COMM(comm,comm_ptr,comm_ptr,myname);

    dtype_ptr = MPIR_GET_DTYPE_PTR(datatype);
    MPIR_TEST_DTYPE(datatype,dtype_ptr,comm_ptr,myname);

    /* Check for invalid arguments */
    if (MPIR_TEST_COUNT(comm,count) ||
    MPIR_TEST_ALIAS(sendbuf,recvbuf))
     return MPIR_ERROR(comm_ptr, mpi_errno, myname );

    MPIR_ERROR_PUSH(comm_ptr);
    /* Test for intercommunicator is done when collops is assigned */

    if(datatype == MPI_LOGICAL)
    {
     int *ps= (int *)sendbuf; /* type cast a void pointer       */
     int *pr;                 /* place declaration here         */
     if(*ps != 0) *ps=1;      /* to use gcc's Allreduce .true.=> 1 */
     sendbuf= ps;             /* ps cannot be put to Allreduce  */

     mpi_errno = comm_ptr->collops->Allreduce(sendbuf, recvbuf, count,
                                      dtype_ptr, op, comm_ptr );
     pr= (int *)recvbuf;
     if(*pr != 0) *pr=-1;     /* back to pgi: .true.=> -1 */
     recvbuf= pr;}
    else
     mpi_errno = comm_ptr->collops->Allreduce(sendbuf, recvbuf, count,
                                      dtype_ptr, op, comm_ptr );
    MPIR_ERROR_POP(comm_ptr);
    TR_POP;
    MPIR_RETURN(comm_ptr,mpi_errno,myname);
}
```

**Table 1.**

Timing of different communication methods using the density-functional ab initio molecular dynamics code Siesta v1.3 [17] for one SCF cycle of a 181 atom system. Four Pentium 4 (3.0GHz), Gigabit Ethernet NIC (3Com996), and PGI Fortran pgf90 are used. MPICH is Argonne National Lab's MPI [20] via TCP/IP, and MPI/GAMMA [10] is via non TCP/IP; "FC on" stands for the use of the flow control during data transmission. For comparison, the last column shows the timing with four typical RISC processors - IBM Power 4 (Regatta, 1.5GHz).

|  |  | Wallclock a) | CPU time b) | (a)-(b) | (a) / (b) |
|---|---|---|---|---|---|
| MPICH TCP/IP |  | 93 sec | 67 sec | 26 sec | 1.39 |
| MPI/GAMMA | FC on | 66 sec | 66 sec | 0.1sec | 1.00 |
|  | FC off | 115 sec | 98 sec | 17 sec | 1.17 |
| RISC machine (1.5GHz) |  | 59 sec | 59 sec | 0.1 sec | 1.00 |

**Figure Captions**

**Figure 1:**
A dual network system is recommended for the Beowulf cluster machine when the GAMMA communication system is adopted. The first network should be a gigabit Ethernet for the GAMMA data communications, and the second one is TCP/IP network either in fast or gigabit Ethernet. The latter is required for the NFS file system (execution binaries are referred to through this network) and for administration purposes.

**Figure 2:**
The relation between transmitted data size (in Bytes) and the data transmission speed (in Mbits/s) is shown for the GAMMA communications. For this measurement, Pentium 4 processors (3.0GHz), and 3Com996 NIC (gigabit Ethernet) are used. The latency at zero data limit is 15micro sec, and the maximum data transmission speed is 706Mbits/s, which is about 70% of the gigabit Ethernet.

**Figure 3:**
Scalability of the density-functional first-principle (quantum mechanical) molecular dynamics code Siesta [16] is shown under the GAMMA and TCP/IP communication systems in filled and open circles, respectively. The relative computational speed is shown in comparison with the uni-processor case. Here, Pentium 4 (3.0GHz) processors, 3Com996 NIC, and Portland's Fortran compiler pgf90 are used. Almost linear scalability is obtained up to four processors for the Siesta run with 181 atoms, and the increase rate in the computational speed is larger for the GAMMA communications than for TCP/IP.

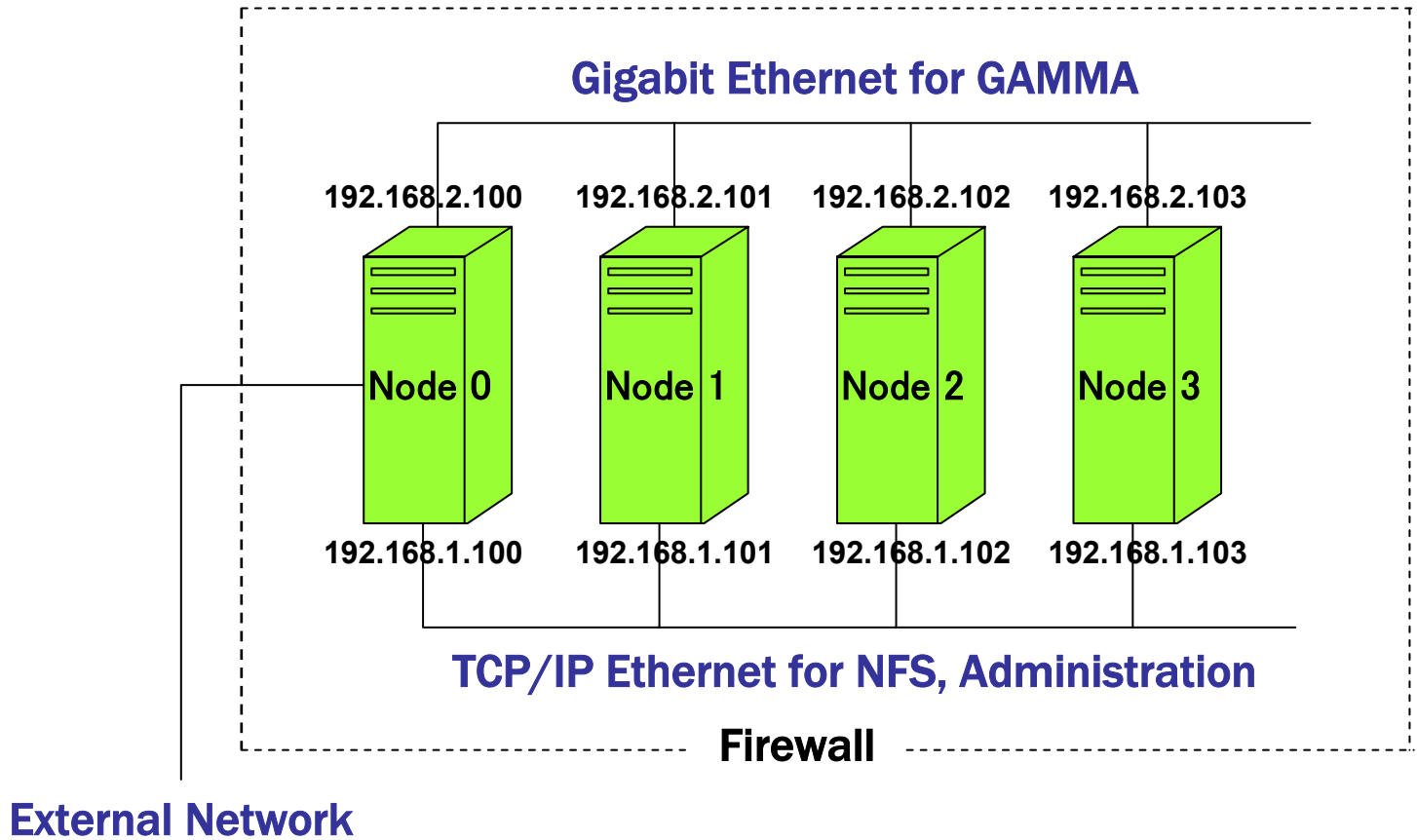

Fig.1  M.Tanaka

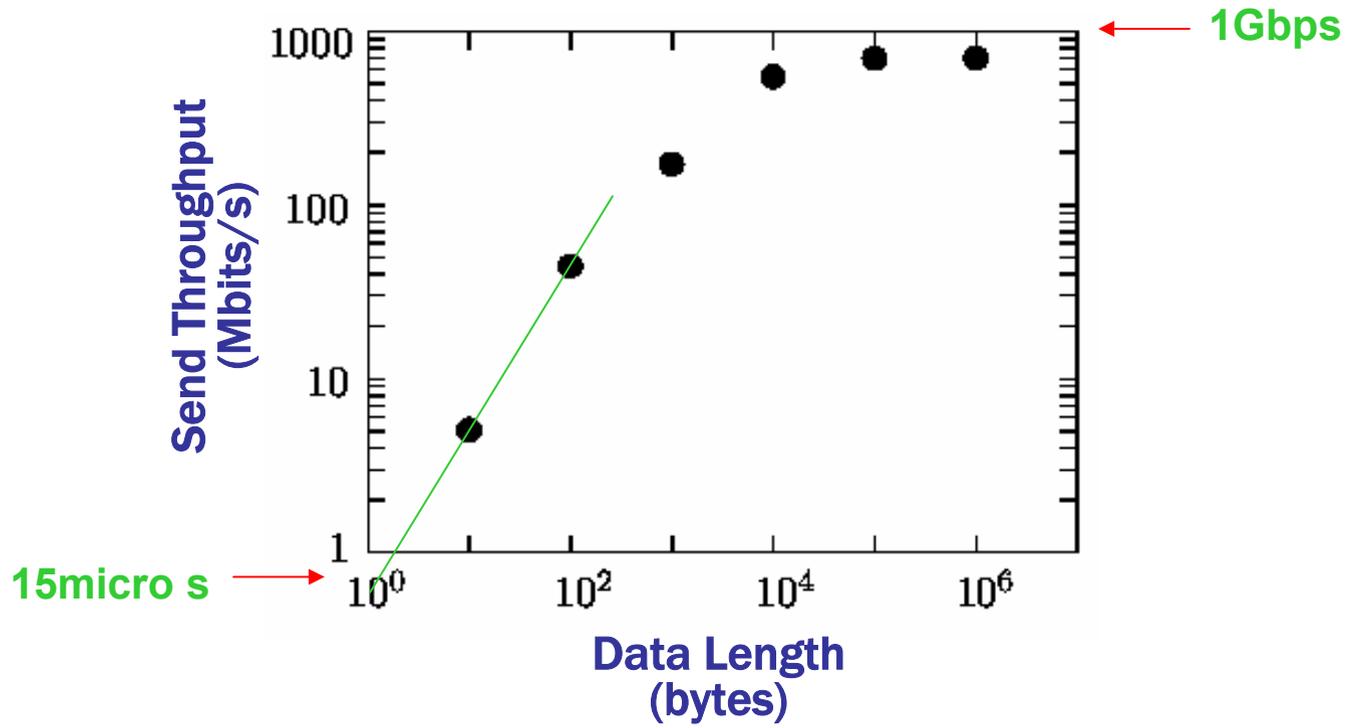

Fig.2   M.Tanaka

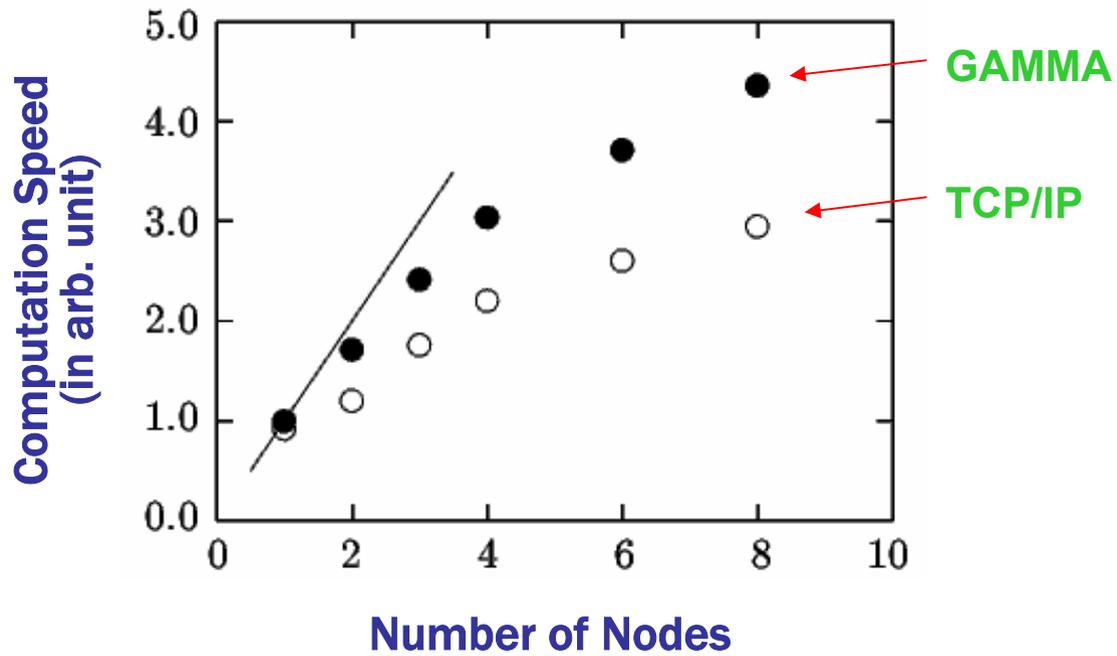

Fig.3   M.Tanaka